\documentstyle[aps,prl,epsf,floats]{revtex}
\begin{document}
\twocolumn[\hsize\textwidth\columnwidth\hsize\csname@twocolumnfalse%
\endcsname
\title{Double-layer quantum Hall antiferromagnetism
at filling fraction $\nu =2/$(odd integer)}
\author{S. Das Sarma${}^{1}$, Subir Sachdev${}^{2}$, and Lian
Zheng${}^{1}$ }
\address{${}^{1}$Department of Physics,
University of Maryland, College Park, MD
20742-4111\\
${}^{2}$Department of Physics, Yale University, P.O. Box 208120,
New Haven, CT 06520-8120}
\date{January 14, 1997}
\maketitle

\begin{abstract}
A low energy action for double-layer quantum Hall systems
at filling fractions $\nu = 2/m$ ($m$ an odd integer) is introduced.
Interlayer antiferromagnetic exchange induces a phase with
canted spin order, and also a spin-singlet phase. Universal properties
of zero and finite temperature transitions are obtained.
We compute the critical temperature at which the canted order
vanishes in a Kosterlitz-Thouless transition.
Implications for recent light scattering experiments at $\nu = 2$
are noted.
\end{abstract}
]
There has been much recent work on
double-layer quantum Hall systems, the majority of which has
focussed on the case where the electron tunneling rate between the
two layers is small~\cite{wenzee,girvin1,girvin2}. Then, the
electron layer
index plays the role of a pseudospin, and for the case where the total
filling factor $\nu = 1/m$ ($m$ an odd integer), very interesting
new physics arises from
long-range correlations in the pseudospin
orientation.
However, the tunneling term acts
like a `magnetic field' in pseudospin space, and so
spontaneous long-range pseudospin
order, and the associated finite temperature ($T$) phase transition,
is only possible
when the tunneling is vanishingly small~\cite{note}.

Stimulated by recent light scattering experiments~\cite{pinczuk} at
$\nu=2$,  we present here a general low energy theory for
double-layer systems at filling $\nu = 2/m$ in the presence
of moderate interlayer tunneling.
We find a rich phase
diagram with interesting transitions both at
$T=0$ and $T> 0$. In contrast to the phases at $\nu = 1/m$,
which are driven by ordering in pseudospin space,
the phases at $\nu = 2/m$ are associated with ordering in
the physical electronic spin space.
Consequently, our order parameters are defined even in the presence
of interlayer tunneling; indeed, moderate
interlayer tunneling is required to stabilize some of our phases.
We will use our results to
interpret recent experiments\cite{pinczuk}, and argue
that they show indirect evidence for our $T>0$ phase transition.

It is useful to begin discussion of the physics at $\nu = 2/m$
by considering the case where the layer separation, $d$, is much larger
than the magnetic length, $l_o$. Then the two layers (labeled 1,2)
are approximately decoupled, and each separately has filling fraction
$\nu_1 = \nu_2 = 1/m$. Their ground states
will be
the familiar Laughlin states for $m>1$, or a fully filled lowest Landau
level
at $m=1$, both of which have a
large energy gap to
all charged excitations~\cite{note2}.
These states are also fully spin polarized and there is
significant intralayer ferromagnetic exchange
\cite{girvin1,girvin2,skyrmion}.
The low-lying
excitations in each layer are spin waves which have a small excitation
gap given precisely
by the Zeeman energy $g \mu_B H$ (the gyromagnetic ratio $g$ and the
Bohr magneton
$\mu_B$ will henceforth be absorbed by a rescaling of the magnetic field
$H$).
For small $g$, a complete description~\cite{girvin2,nr} of the low
energy
excitations of each layer
can be given in terms of an action for unit vector fields
$\vec{n}_{1,2}$ (
$\vec{n}_{1,2}^2 = 1$) representing the orientation of the ferromagnetic
orders.
Spin waves are small fluctuations of $\vec{n}_{1,2}$ about an ordered
state,
while charged quasiparticles are skyrmion~\cite{girvin2,skyrmion}
textures of $\vec{n}_{1,2}$.

Now reduce the value of $d$ and couple $\vec{n}_1$ and $\vec{n}_2$.
The simplest allowed coupling between them is an {\em
antiferromagnetic\/} exchange interaction. These considerations
lead to the following imaginary-time ($\tau$) effective action
(in units with $\hbar=k_B = 1$)
\begin{eqnarray}
&& {\cal S}_0 = \int d^2 x \int_0^{1/T}  d \tau
\left( {\cal L}_F [\vec{n}_1] + {\cal L}_F [\vec{n}_2 ] + J \vec{n}_1
\cdot \vec{n}_2
\right) \nonumber \\
&& {\cal L}_F[\vec{n}] \equiv iM_0 \vec{A}(\vec{n}) \cdot
\partial_{\tau} \vec{n}
+ \frac{\rho_s^0}{2} \left( \nabla_x \vec{n} \right)^2 - M_0 \vec{H}
\cdot \vec{n}
\label{action1}
\end{eqnarray}
The intralayer ferromagnetic spin
correlations~\cite{girvin2,skyrmion,nr}
are controlled by ${\cal L}_F$:
$M_0 = 1/4 \pi m l_o^2$ is the magnetization density per layer,
$\rho_s^0$ is the spin stiffness of each layer when they are well
separated
(for $m=1$, we have~\cite{kallin}
$\rho_s^0 = e^2/(16 \sqrt{2\pi} \epsilon \l_o$) and $\vec{A}$ accounts
for the Berry phase accumulated under time evolution of the spins
($\epsilon_{ijk} \partial A_k (n) / \partial n_j = n_i$). The
interlayer antiferromagnetic
correlations are induced by the positive coupling $J \sim M_0
\Delta^2_{\rm sas}/ U$
where
$\Delta_{\rm sas}$
is the tunneling matrix element between the layers, and $U \sim e^2
/ \epsilon \l_o$
is the Coulomb interaction energy.

Some potentially important terms have been omitted from ${\cal S}_0$
and our analytic computations: the Hopf term which endows the skyrmions
with fractional statistics,
and the long-range Coulomb
interaction between the skyrmions. We believe this is permissible
because of the charge
gap noted earlier. Further~\cite{scs}, as the layers are
antiferromagnetically correlated,
skyrmions in one layer will be correlated with anti-skyrmions in the
other, and this
neutralizes the leading contribution of both terms. This latter argument
should continue to hold
even if the charge gap were to vanish at a quantum critical point.
Note that no new term is necessary to induce charge transfer
between the layers:
a hedgehog/anti-hedgehog pair in the two layers corresponds to an event
transferring
skyrmion number between them. Such spacetime singularities are absent in
the
strict continuum limit but appear when a short-distance regularization
is introduced. Finally, for $m>1$ and larger $g$, the spin 0 Laughlin
quasiparticles become the lowest energy charged excitations, but these
can be neglected for similar reasons.

Now we parameterize
\begin{equation}
\vec{n}_i = (-1)^i (1 - \vec{L}^2 )^{1/2} \vec{n} + \vec{L}
\label{nl}
\end{equation}
where the constraints $\vec{n}_{1,2}^2 = 1$ are now replaced by
$\vec{n}^2 = 1$
and $\vec{L} \cdot \vec{n} = 0$. Because the layers are
antiferromagnetically correlated
we expect that $\vec{L}$ will not be too large. We insert
(\ref{nl}) into ({\ref{action1}), expand to quadratic order in
$\vec{L}$, and then
integrate out the $\vec{L}$ degrees of freedom. This yields the
following
effective action for the antiferromagnetic order parameter $\vec{n}$
\begin{displaymath}
{\cal S}_1 = \frac{c}{2t} \int d^2 x \int_0^{1/T}  d \tau
\left[ (\nabla_x \vec{n} )^2 + \frac{1}{c^2} \left( \frac{\partial
\vec{n}}{\partial \tau}
- i \vec{H} \times \vec{n} \right)^2 \right]
\end{displaymath}
where $t = (J / 2 \rho_s^0 M_0^2)^{1/2}$ and $c= ( 2 \rho_s^0
J/M_0^2)^{1/2}$.
This is precisely the action of the 2+1 dimensional quantum $O(3)$
non-linear
sigma model in a field $H$ coupling to the conserved global $O(3)$
charge. It is expected
to apply to double-layer quantum Hall systems with $\nu = 2/m$
at length
scales larger
than $\Lambda^{-1} \sim l_o$.

For the special case $m=1$,
we will sharpen the quantitative theoretical predictions
of ${\cal S}_1$ by a Hartree-Fock (HF) analysis of a realistic,
microscopic double-layer Hamiltonian~\cite{zrd1}.
The HF theory will be used to compute {\em renormalized\/}
$T=0$ energy scales
which completely specify the correlators of ${\cal S}_1$ at low $T$:
in this manner we determine observables of the system with no free
parameters (for $m>1$ these energy scales remain as phenomenological
parameters). We use the Hamiltonian ${\cal H} =
{\cal H}_0 + {\cal H}_{\rm I}$ with
\begin{displaymath}
{\cal H}_0=-{\Delta_{\rm sas}\over2}\sum_{\alpha\sigma}\ \left(
C^\dagger_{1\alpha\sigma}
C_{2\alpha\sigma}+h.c.\right)-{H\over2}\sum_{i\alpha\sigma}{\sigma}
C^\dagger_{i\alpha\sigma}C_{i\alpha\sigma},
\end{displaymath}
where $C_{i\alpha\sigma}$ annihilates an electron in the lowest Landau
level
in layer $i$ ($i=1,2$) with spin $\sigma$ ($\sigma = \pm 1$)
in the $z$ direction (we assume $\vec{H} = (0,0,H)$) and
with intraLandau level index $\alpha$.
Interlayer tunneling induces the symmetric-antisymmetric energy
separation
$\Delta_{\rm sas}$.
The Coulomb interaction part of ${\cal H}$ is
\begin{eqnarray}
{\cal H}_{\rm I}=&&{1\over2}\sum_{\sigma_1\sigma_2}\sum_{ij}
\sum_{\alpha_1\alpha_2}{1\over\Omega}\sum_{\bf q}
V_{ij}(q)e^{-q^2l_o^2/2}e^{iq_x(\alpha_1-\alpha_2)l_o^2}
\nonumber \\
&&\times C^\dagger_{i\alpha_1+q_y\sigma_1}C^\dagger_{j\alpha_2\sigma_2}
C_{j\alpha_2+q_y\sigma_2}C_{i\alpha_1\sigma_1},
\label{equ:hi}
\end{eqnarray}
where $q$ is a wavevector, $\Omega$ is the area of the sample,
and the interaction potentials are
$V_{ij}={2\pi e^2/\epsilon q}$ for $i=j$ and
$V_{ij}={(2\pi e^2/\epsilon q})e^{-qd}$ for $i\ne j$.

The $T=0$ phase diagram~\cite{ss1,annals} for the action
${\cal S}_1$ is shown in Fig. \ref{fig1}.
For $\nu=2$ a topologically identical phase diagram
is obtained in the HF computation~\cite{zrd1}, and is
shown as an inset.
The quantum phase transitions between these phases are continuous
and are accompanied by the softening of the intersubband spin density
excitations.
The phases are described below:

\begin{figure}
\epsfxsize=3.5in
\centerline{\epsffile{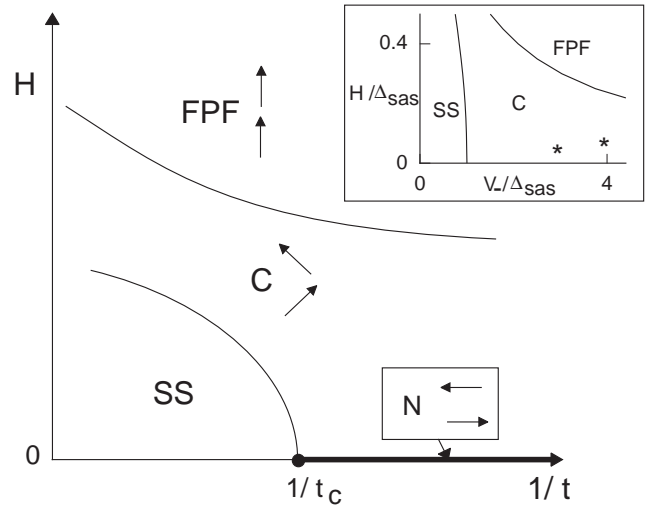}}
\caption{$T=0$ phase diagram of ${\cal S}_1$.
The phases are pictorially
represented
by the orientation of the spins in the two layers, with $H$ pointing
vertically upwards;
$C$ ($N$) has a
broken $O(2)$ ($O(3)$)
spin symmetry. There is a Kosterlitz-Thouless transition at $T=T_c > 0$
in $C$.
The inset shows the phase diagram obtained from a microscopic
HF calculation at $\nu=2$ ($m=1$),
where the asterisks represent the two experimental samples of
ref.~\protect\onlinecite{pinczuk}.
We argue in the text that the HF theory overestimates the
stability of the $C$ phase,
and that experiments suggest that the
actual $SS$ region encloses the left
sample in the inset.
}
\label{fig1}
\end{figure}

\noindent ({I\/}) Fully Polarized Ferromagnet ($FPF$):
In ${\cal S}_1$ this is present for $H > \sim c t \Lambda^2$.
This phase has $\langle n_{1z}
\rangle =
\langle n_{2z} \rangle = 1$. It is continuously connected to the large
$d
$ limit
discussed earlier.\\
({II\/}) Canted ($C$):
We now have $\langle n_{1z} \rangle = \langle n_{2z} \rangle \neq 0$,
and, for example, $\langle n_{1x} \rangle = - \langle n_{2x} \rangle
\neq
0$. This phase has a
broken spin rotational $O(2)$ symmetry in the $x-y$ plane.
For $m=1$, the HF phase boundary between the $FPF$ and $C$ phases is
$V_-/\Delta_{\rm sas} =(\Delta_{\rm sas}/H)[1-(H/\Delta_{\rm sas})^2]$,
and that
between the $C$ and  $SS$ phases is
$V_-/\Delta_{\rm sas}=1-(H/\Delta_{\rm sas})^2$,
where $V_\pm=
{1\over\Omega}\sum_{\bf q}e^{-q^2l_o^2/2}[V_{11}(q)\pm V_{12}(q)]$;
a wavefunction for the $C$ phase is obtained by
the standard HF methods. For $m>1$
a caricature of the wavefunction is two separate Laughlin states at
$\nu_1 =
 \nu_2 = 1/m$
but polarized in the orientations shown.\\
({III\/}) Neel ($N$):
This is the limiting case of $C$ with $\langle n_{1z} \rangle = \langle
n_{2z} \rangle = 0$
achieved at $H=0$. Now an $O(3)$ spin rotation symmetry is broken.\\
({IV\/}) Spin Singlet ($SS$):
This corresponds to the quantum disordered phase of the $O(3)$ sigma
model.
The
ground state is a
spin-singlet and is therefore unaffected by $H$: its wavefunction is the
same as that at $H=0$.
For $m=1$, in the independent electron HF picture, the electrons fill
the
layer-symmetric
subband, with spin-up and spin-down levels equally populated.
However,
it is well-known that HF theory overestimates the
energy of a {\em nonmagnetic\/} phase like $SS$ because correlations
between opposite spin electrons, important for reducing the Coulomb
energy,
are now absent.
It is likely, therefore, that the {\it real} $SS$ phase is
stable over a larger parameter region than that
in our HF approximation, but the topology of the HF phase
diagram in the  inset of Fig.~\ref{fig1}
(which is identical to that for ${\cal S}_1$)
should be correct.
To build in charge correlations,
one can use an approach similar to the Heitler-London
picture of the hydrogen molecule, and consider pairs of electrons
with their charge localized in opposite layers, while their spins
form singlet bonds. Indeed, such a charge-localized picture
was behind our introduction of the actions
${\cal S}_{0,1}$. In such an approach, an alternative wavefunction
for the $SS$ phase (valid for $m=1$ {\em and\/} $m>1$) can be obtained
in the $J \rightarrow \infty$ limit:
 pairs of electrons in opposite layers bind to form spin singlet, charge
$2e$
bosons, which then condense into a boson Laughlin state at filling
fraction
 $1/2m$, as
demanded by the strength of the magnetic flux.

It is worth noting explicitly here that the HF computations at
$m=1$ allow us to assert that all the different phases of
${\cal S}_1$ are the ground states in realistic
parameter regimes. For $m>1$, it remains
an open question as to whether the phases of ${\cal S}_1$ other than
$FPF$ are accessible, although we consider it a likely possibility that
at least $C$ will exist.

We now turn to the physics at $T>0$. Only the $N$ and $C$ ground states
have a broken spin rotation symmetry; the $O(3)$ symmetry of the former
implies that the symmetry is restored at any $T>0$, while the
$O(2)$ symmetry of the latter implies a Kosterlitz-Thouless phase
transition at a $T=T_c > 0$. We may characterize the order parameter
fluctuations in both phases by a $T=0$ spin stiffness $\rho_s (H)$ such
that the energy cost of rotations of the order parameter by a
slowly varying angle $\phi(r)$ is
$E_\phi=(\rho_s (H)/2)\int\ d^2r |\bigtriangledown\phi({\bf r})|^2$.
A crude estimate \cite{girvin2,zrd1}
 of $T_c$ is $T_c \approx \rho_s (H)$, although
this must fail as $H \rightarrow 0$. In the latter limit it
is possible to obtain an exact leading asymptotic result~\cite{ss1}
$T_c = 2 \pi \rho_s (0)/\ln(\rho_s (0)/H)$ for $\ln (\rho_s (0)/H) \gg
1$.
For $m=1$ we computed $\rho_s (H)$ in the HF calculation and
the results are shown in Fig.~\ref{fig2}. We see that the $T_c$
estimates
are well in the experimentally accessible regimes for
typical GaAs-based semiconductor samples.
We emphasize that the Kosterlitz-Thouless transition at $T_c$
is present even in the presence of interlayer tunneling,
unlike the case for the pseudospin transition \cite{girvin1,girvin2}
at $\nu = 1/m$.

\begin{figure}
\epsfxsize=2.8in
\centerline{\epsffile{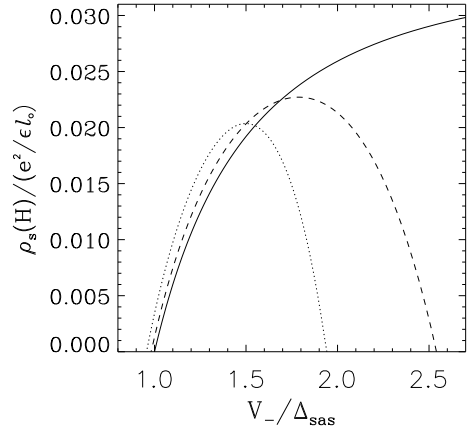}}
\caption{Ground state spin stiffness $\rho_s (H)$
 of a $\nu=2$ double-layer system
obtained from the microscopic HF
calculations. It is non-zero only in the $N$ and $C$ phases.
The solid line has $H=0$, the dashed line $H=0.05 e^2/\epsilon l_o$,
and the dotted line $H=0.08 e^2/\epsilon l_o$.
The layer separation is
$d=1.0l_o$ and we have also included corrections from
finite layer thickness $d_w=0.8 l_o$.
In this figure,  
$V_-$ is fixed to be $0.36 e^2/\epsilon l_o$
for the given values of $d$ and $d_w$,
and $\rho_s(H)$ is shown as 
a function of $\Delta_{\rm sas}$ for several values of $H$.
(In typical GaAs-based samples, $e^2/\epsilon l_o$ is on the order
of $50-100K$, which gives $\rho_s\sim 1-2K$.)
}
\label{fig2}
\end{figure}

A more precise approach to the $T>0$ properties
is to expand in the deviation from the
$H=T=0$ quantum critical point between the $N$ and $SS$ phases
at $t = t_c \sim \Lambda^{-1}$. This quantum critical point is described
by a
renormalizable quantum field theory (with upper-critical spatial
dimension
$d=3$), and so all thermodynamic properties are
universal functions
of energy scales characterizing `relevant' perturbations from this
critical point;
corrections due to irrelevant operators require additional energy scales
and will
be neglected here.
Two of the relevant energy
scales are the `bare' couplings $T$ and $H$ (there is no
renormalization of the scale
of $H$ because it couples to a conserved charge~\cite{ss1}) and a
third (the last) measures
deviation of $t$ from $t_c$. For $t > t_c$ we choose~\cite{ss3} this
energy scale to be
$\Delta$, the energy gap of the $SS$ state at $T=H=0$, while for $t<t_c
$
we choose
the renormalized spin stiffness $\rho_s (0)$ of the $N$ state also at
$T=H=0$. As $t$ approaches $t_c$ we have
$\Delta \sim (t-t_c)^{\nu}$, while $\rho_s (0) \sim (t_c -
t)^{(d-1)\nu}$,
where $\nu$ is the correlation length exponent of the {\em classical}
three-dimensional $O(3)$
ferromagnet.
For $m=1$, the microscopic HF calculation gives
$\Delta=\Delta_{\rm sas}\sqrt{1-V_- / \Delta_{\rm sas}}$,
and $\rho_s (0) =((1-(\Delta_{\rm sas}/V_-)^2)/8\pi\Omega)
\sum_{\bf q} (ql_o)^2e^{-q^2l_o^2/2}
V_{11}(q)$. Notice that these are consistent with the
mean field exponent $\nu=1/2$ in the upper-critical dimension
$d=3$.

One of our main results, which follows from the considerations above,
is that the critical
temperature $T_c$ at which
the ordering of the $C$ phase disappears obeys, for $t>t_c$,
\begin{equation}
T_c = H \Psi_> ( \Delta / H).
\end{equation}
Here $\Psi_> (u)$ is a universal function of $u$ with no arbitrary scale
factors,
and obeys the exact relation $\Psi_> (u \geq 1) = 0$
(because~\cite{zrd1,ss1}
the $T=0$ boundary
of the $SS$ phase is given precisely by the condition $\Delta=H$).
A similar scaling form holds for $t < t_c$ with $T_c = H
\Psi_< (\rho_s (0)/H)$.
We computed the functions $\Psi_{>,<}$ in an expansion in $\varepsilon =
3-d$ using recently developed
methods~\cite{ss2} and found
to leading order
\begin{equation}
\Psi_> (u) = [33 (1-u^2)/(10 \pi^2 \varepsilon)]^{1/2};
\end{equation}
the structure of the subleading terms is quite complicated and is
similar to that discussed
elsewhere~\cite{ss2}. This result is valid for all $u$, except for $u$
very close to 1;
in that case we find, by a mapping to the dilute Bose gas problem,
the exact asymptotic result~\cite{ss1,popov} $\Psi_> (u \rightarrow 1) =
y
\ln (1/y)/(4 \ln \ln (1/y))$
with $y=1-u$, which holds for $\ln (1/y) \gg 1$. For $t<t_c$ the
$\varepsilon$
expansion holds for $\Psi_< ( u/\sqrt{\varepsilon})$ and we obtained
\begin{equation}
\Psi_< (u/\sqrt{\varepsilon}) = [(33+ 3 u^2)/(10 \pi^2
\varepsilon)]^{1/2}
\label{e2}
\end{equation}
Again this result is valid for all $u$, but now fails for $u \rightarrow
\infty$
(which is $H \rightarrow 0$).
While $T_c (H=0) > 0$ for all $\varepsilon < 1$, we noted earlier
that $T_c (H=0) = 0$ for $\varepsilon = 1$; the latter
property will not appear at any order in
the $\varepsilon$ expansion. Using results special
to $d=2$
for $H \rightarrow 0$ discussed earlier, we have instead the exact
asymptotic form $\Psi_< ( u \rightarrow \infty ) = 2 \pi u/\ln u$.

We draw attention to a particularly simple and striking
limit of the above results. At $t=t_c$ we have $T_c = {\cal K} H$ where
${\cal K}
= \Psi_> (0) = \Psi_< (0)$ is a universal number. Further, we do not
expect
any large or singular variation in $T_c$ if $t$ is close to but not
exactly $t_c$.
As both $H$ and $T_c$ are directly measurable energies, this
relationship is
amenable to a direct
experimental test. On the theoretical side, while at present there is
only the leading term
in a $\varepsilon$ expansion for the value of ${\cal K}$, it should
be possible to obtain a reasonably precise result using quantum Monte
Carlo simulations
of double-layer
lattice spin systems~\cite{scs,double}, which have been limited to $H=0$
so far. Universality implies that these lattice models will have
the same value of ${\cal K}$ as the quantum Hall system, and it appears
to us
that the simulation for $H \neq 0$ should also be free of
the fermion sign problems.

We have also obtained results in the $\varepsilon$ expansion
for the crossovers of the dynamic spin susceptibility at frequency
$\omega$
as universal functions of the energy ratios $\omega/T$, $H/T$ and
$\Delta/T$ ($\rho_s (0)/T$): the methods are similar to those
of ref. \onlinecite{ss2}, and results will be presented elsewhere.

We turn now to a comparison with recent light scattering
experiments \cite{pinczuk}. The high density sample
(the right sample in the inset of Fig~\ref{fig1})
shows `mode-softening' consistent with a $T>0$ phase transition
which we identify with that above our $C$ phase.
Using input parameters from the HF calculation in (\ref{e2}),
we obtain the prediction of $T_c \sim 0.5 K$, to be compared
with the experimental value $T_c \sim 0.52 K$: the good agreement
must be considered fortuitous until the accuracy of the $\varepsilon$
expansion is better understood. The same sample also shows marked $T$
dependence
for $T> T_c$ in the light scattering spectrum
 at $\omega$ of order or greater than $T$:
a natural explanation for this could be a crossover into the `high-$T$'
region
above the quantum-critical point~\cite{ss2}.
The lower density sample
(the left sample in the inset of Fig~\ref{fig1})
shows no mode softening
and little $T$ dependence
in the light scattering spectrum: we suggest that this sample is in the
$SS$ phase.
The HF computation puts this sample in the $C$ phase,
but as we discussed earlier,
this could be in error because the HF theory overestimates the
stability of the $C$ phase.

Finally we note that we expect similar considerations to apply to all
double-layer
quantum Hall systems with $\nu = 2 \nu_1$ where a single layer at
filling $\nu_1$
forms a fully polarized quantum Hall state with a charge gap.

This work was supported
by the U.S.-ONR (S.D.S. and L.Z.)
and by NSF
Grant No DMR 96--23181 (S.S.)


\begin{references}
\bibitem{note} Non-zero $T$ transitions are also possible for a tilted 
magnetic field;
see ref~\protect\onlinecite{girvin1}.
\bibitem{note2} This is a crucial difference from the
$\nu=1/m$ case where two well separated layers cannot form such Laughlin
states.
At $\nu=2/m$ the charge gaps suppress interlayer charge fluctuations,
and consequently the pseudospin order parameter (which is canonically 
conjugate to the
interlayer charge difference) has zero stiffness, and can be neglected.
\end{references}
\end{document}